\documentclass[a4paper,12pt]{revtex4}
\usepackage{graphicx}
\def\beq{\begin{equation}}
\def\eeq{\end{equation}}
\def\beqn{ \begin{eqnarray} }
\def\bar{ \begin{array} }
\def\bdm{ \begin{displaymath}}
\def\esl{ \end{slide}}
\def\eeqn{ \end{eqnarray} }
\def\ear{ \end{array} } 
\def\edm{ \end{displaymath}}
\def\s1s2{{ \mathbf{\sigma_1} \cdot \mathbf{\sigma_2} }}
\def\t1t2{{ \mathbf{\tau_1} \cdot \mathbf{\tau_2}  }}

\def\ap{a_p}
\def\ap'{a_p'}
\def\ap1{a_{p_1}}
\def\ap'1{a_{p'_1}}
\def\ap2{a_{p_2}}
\def\ap'2{a_{p'_2}}
\def\ah{a_h}
\def\ah'{a_h'}
\def\ah1{a_{h_1}}
\def\ah'1{a_{h'_1}}
\def\ah2{a_{h_2}}
\def\ah'1{a_{h'_2}}
\def\acp{a\dag_p}
\def\acp'{a\dag_p'}
\def\acp1{a\dag_{p_1}}
\def\acp'1{a\dag_{p'_1}}
\def\acp2{a\dag_{p_2}}
\def\acp'2{a\dag_{p'_2}}
\def\ach{a\dag_h}
\def\ach'{a\dag_h'}
\def\ach1{a\dag_{h_1}}
\def\ach'1{a\dag_{h'_1}}
\def\ach2{a\dag_{h_2}}
\def\ach'1{a\dag_{h'_2}}

\newcommand{\bsigma}{\mbox{\boldmath $\sigma$}}
\newcommand{\btau}{\mbox{\boldmath $\tau$}}

\newcommand{\car}{$^{12}$C~}
\newcommand{\oxy}{$^{16}$O~}
\newcommand{\caI}{$^{40}$Ca~}
\newcommand{\caII}{$^{48}$Ca~}
\newcommand{\zr}{$^{90}$Zr~}
\newcommand{\pb}{$^{208}$Pb~}
\begin{document}

\title{Effective nucleon-nucleon interaction and low-lying 
nuclear magnetic states}

\author{C. Maieron$^{1}$
\footnote{Email address: chiara.maieron@le.infn.it}, 
V. De Donno$^{1}$, G Co'$^1$, 
M. Anguiano$^{2}$, 
A. M. Lallena$^{2}$ and  M. Moreno Torres$^{2}$}

\affiliation{$^{1}$ Dipartimento di Fisica, Universit\`a del Salento and INFN,
  sezione di Lecce,
Lecce, I-73100, Italy\\
$^{2}$ Departamento de Fis\'ica At\'omica, Molecular y Nuclear,
Universidad de Granada,
Granada, E-18071, Spain}

\begin{abstract}
 We present a calculation of  low energy magnetic states of
          doubly-closed-shell nuclei. Our results have been obtained
          within the random phase  approximation using different
          nucleon-nucleon interactions, having zero- or finite-range
          and including a possible contribution in the tensor channel.

\end{abstract}
\maketitle

\section{Introduction}
\label{sec:intro}

In the last thirty years electron-nucleus scattering experiments have 
produced a large amount of high precision data, which impose severe
constraints on nuclear models and effective theories, such as the
Random Phase Approximation (RPA). In particular the description of
low energy excited states within the RPA is known to be very sensitive 
to the details of the effective nucleon-nucleon (NN) interaction
used in the calculations.

We present here a selection of results
from a systematic study of 
the low energy spectra of several
doubly-closed-shell nuclei we have made within the RPA theory
\cite{don08t}. We have focused, in particular, on the unnatural parity states, 
which are sensitive to the spin,
spin-isospin and tensor channels of the residual NN interaction
\cite{don09}.

In the first step of our project 
we have employed a purely phenomenological approach, 
using a single-particle mean-field basis  generated
by a  Woods-Saxon well, and constructing  phenomenological residual
NN interactions which reproduce some selected nuclear states.
In order to study the sensitivity of our results to the details of the
residual interaction, we have used four different NN interactions,
which have zero- and finite-range and may include contributions in the
tensor channels. We have found some states which are very
sensitive
to the details of the interaction, and in general we 
have obtained a satisfactory description of the low energy states of the
nuclei under consideration. We can thus consider our phenomenological
approach
as an ``optimal'' RPA approach in terms of comparison with the
experimental data. 

In our study we have then used the RPA amplitudes
$X$ and $Y$ to calculate electron scattering response functions.
As a further independent extension
of  this approach, we have also considered the computation 
of neutrino cross sections, for which an example will be shown 
in the following. 

We have then proceeded to the second step of our study, performing RPA
calculations within a self-consistent approach, where the
single-particle
mean-field 
basis is obtained by means of a Hartree-Fock calculation which uses the same
effective
NN interaction used in the RPA calculation. In particular we have used
the Gogny D1 finite-range interaction~\cite{gog75,bla77,dec80}, 
finding, in this case,
remarkable disagreement with the experimental spectra.   

\section{Formalism}
\begin{figure}[t]
\begin{center}
\includegraphics[width=3.8in, angle=90] {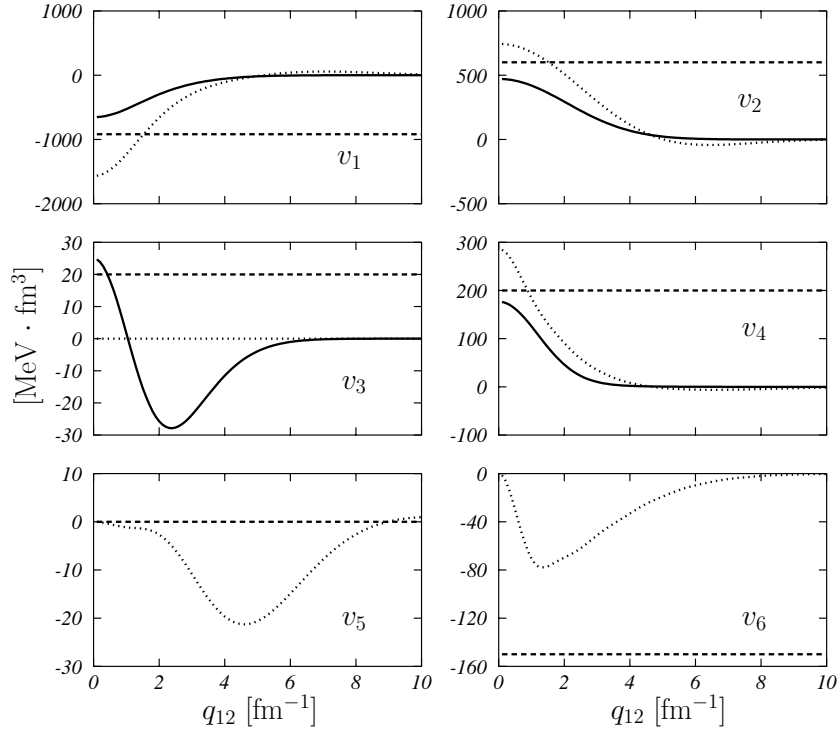} 
\end{center}
\caption{\small Effective 
 NN interactions as  functions
 of the relative momentum. 
The dashed lines represent the LMtt interaction, the 
dotted the FRtt and the solid the Gogny D1. The central channels 
$v_i$ ($i$=1-4) of the LMtt and FRtt interactions
are identical to those of  LM and FR, respectively. 
}
\label{fig:NNint}
\end{figure}
%

The inputs required by RPA calculations are  a set of
single-particle
energies and wave functions and a residual NN interaction.

In the purely phenomenological calculations, for the former we have used 
the single-particle basis generated by a Woods-Saxon well, whose
parameters are taken from the literature~\cite{ari07}.
For the latter, we have considered a generic residual interaction written as
\beqn
V_{\rm eff}(1,2)&=&
       v_1(r_{12}) \, 
  + \, v_1^\rho(r_{12})\, \rho^\alpha(r_1,r_2) \nonumber \\
&&+ \, \left[ v_2(r_{12}) \, 
  + \, v_2^\rho(r_{12}) \, \rho^\alpha(r_1,r_2) \right] \, \t1t2\nonumber \\
&&+ \, v_3(r_{12})\, \s1s2 \, 
  + \, v_4(r_{12}) \, \s1s2 \, \t1t2  \nonumber \\
&&+ \, v_5(r_{12}) \, S_{12}(\hat{r}_{12}) \, 
  + \, v_6(r_{12})\, S_{12}(\hat{r}_{12}) \, \t1t2 
\, ,
\label{eq:intr}
\eeqn
where  $\bsigma$ and $\btau$ are the usual spin and 
isospin operators  and $S_{12}$ is the tensor operator. 
As suggested by  past phenomenological  RPA studies~\cite{bla77,spe77} 
we have included density dependent terms in the
the central and isospin channels, with
\beq
\rho(r_1, r_2)={\left[\rho(r_1)\rho(r_2)\right]}^{1/2} 
\,\,,
\label{eq:vdens2}
\eeq
and using $\alpha=1$ in Eq.~(\ref{eq:intr}).

We have considered four different forms of the NN interaction, parametrizing
them
according to the following criteria:
{\it (i)} we have chosen a unique set of parameters for all the nuclei under
  investigation, with the exception of the density dependent terms
which are different for each nucleus;
{\it (ii)}  the density dependent terms have been set
to reproduce the first $2^+$ state in \car and the first $3^-$ states
in the other nuclei;
{\it (iii)} the remaining contributions in the central and isospin channels
  have been chosen to get a reasonable description of the centroid
  energy 
of the
isovector giant dipole resonance;
{\it (iv)} the spin,
spin-isospin and tensor channels 
have been adjusted to describe the low energy (below 8 MeV) magnetic spectrum
of \pb, 
with particular attention to the $12^-$ and $1^+$ states  
and, in addition, taking care that
the energy of the first $4^-$ state of \oxy is reproduced reasonably.
%
\begin{figure}[t]
\begin{center}
\includegraphics[width=3.8in,angle=90]{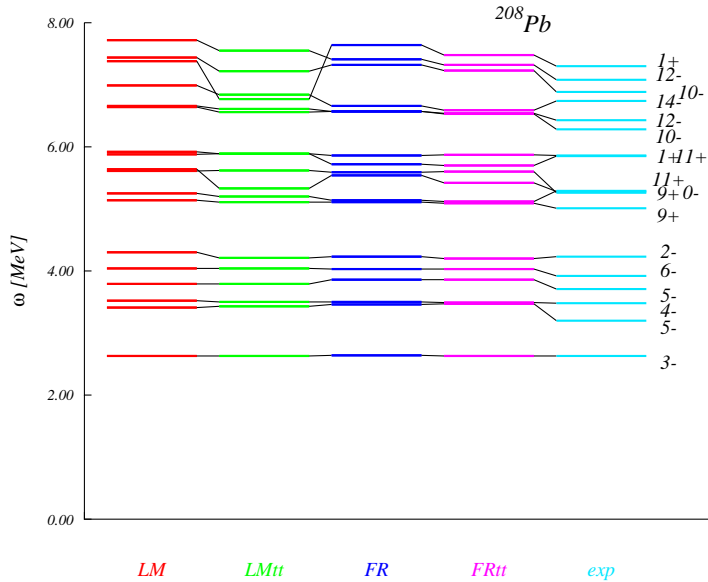}
\end{center}
\vskip -0.5cm
\caption{(Color online) Low energy spectrum of \pb calculated with the four
  interactions
used in this work and compared with the experimental energies~\cite{led78}.} 
\label{fig:pbspec}
\end{figure}
 
Following  a Landau-Migdal approach, we have first considered zero-range
interactions without and with a tensor-isospin channel contribution
(LM and LMtt in the following).

We have then constructed finite-range interactions (FR and FRtt), 
by keeping the long-range
behavior of the Argonne $v_{18}$ potential~\cite{wir95} and substituting its
short-range part with a sum of Gaussians.
We have also used Gaussians to parametrize the density dependent terms of
the interaction and, for the FRtt case, 
we have obtained the tensor channel terms by multiplying 
the corresponding terms of the $v_{18}$ interaction by a correlation function
obtained in variational calculations~\cite{ari07}.

The behavior of the various interactions we have considered is shown in
figure~\ref{fig:NNint} as a function of the relative momentum of the
interacting pair.

In the self-consistent approach we have obtained the single particle basis
by solving  Hartree-Fock equations ~\cite{co98,bau99},
using the same effective NN 
interaction used for the RPA calculations. We have used the
Gogny D1 interaction~\cite{gog75,bla77,dec80}, shown in
figure~\ref{fig:NNint} by the solid lines, which has finite-range
components in the central, isospin, spin and spin-isospin channels and
a zero-range density dependent contribution. 
We have not included the spin-orbit term in the
RPA calculations.

For the self-consistent 
calculations we have included the contributions of both direct and
exchange
matrix elements of the interaction,  whereas for the phenomenological 
approach we have
considered direct terms only, assuming that the effect of the exchange
terms
is effectively included in the choice of the parameters characterizing
the
various NN interactions.

Finally we observe that our calculations have been obtained by
discretizing the continuum. We have checked that our results are
stable
with respect to the parameters characterizing the continuum
discretization and also with respect to the size of the
single-particle configuration space used for each nucleus. 
More details on the role 
of the continuum discretization in self-consistent RPA
calculations can be found in~\cite{don09}. 

\section{Results}

\begin{figure}[t]
\begin{center}
\includegraphics[width=3.5in]{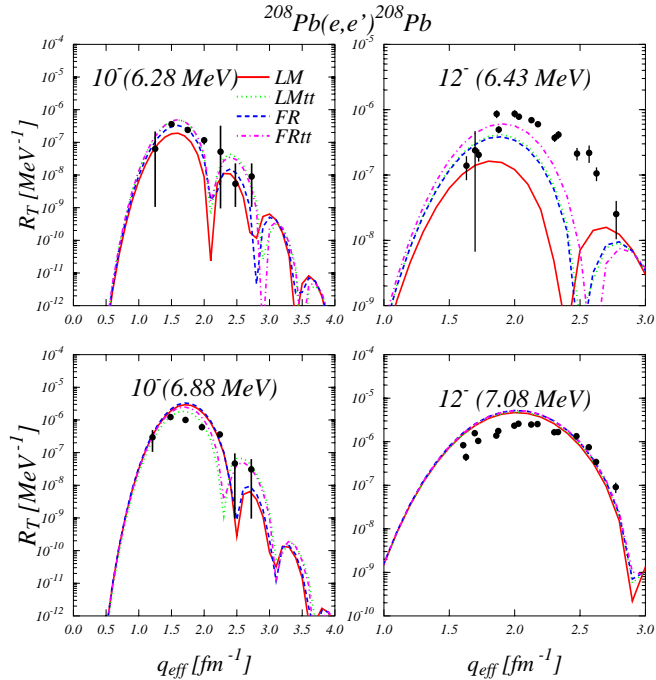}
\end{center}
\vskip -1cm
\caption{(Color online) Right panels: electron scattering transverse responses 
for the first $12^-$ states of \pb, versus the effective momentum
transfer. Different residual interactions are used as indicated.
Left panels: the same for the first $10^-$ states.  
Experimental data from~\cite{lin79}.}
\label{fig:pb12m}
\end{figure}

We have performed systematic calculations of the low energy spectra
of \car, \oxy, \caI, \caII, \zr 
and  \pb. For each state we have also computed
electromagnetic
response functions, and we have compared
them with the available experimental data.
\begin{table}[b]
\begin{center}
\caption{\label{tab:c12}Energies of the low-lying magnetic states of
  \car (in MeV). Experimental energies from 
\cite{led78}.}
{\begin{tabular}{|c|c|c|c|}
\hline
\multicolumn{4}{|c|}{\car}\\
\hline
excitation&D1&FRtt&exp\\
\hline
$1^+$ &11.17 &13.87&12.71\\
$1^+$&7.73 &18.05& 15.11 \\ 
$4^-$&19.16 &17.75& 18.27 \\ 
$4^-$&15.63 &19.49& 19.15 \\ 
\hline
\end{tabular}}
\end{center}
\end{table}

As an example of low energy spectrum, in figure~\ref{fig:pbspec} we show
the case of \pb. Here we present only those experimental states we
have been able to identify with those obtained in our calculations.
We observe that the general agreement is quite good and that,  
except for some cases, the various 
energies have little sensitivity to the details of the interactions.
\begin{figure}[t]
\begin{center}
\includegraphics[width=3.5in]{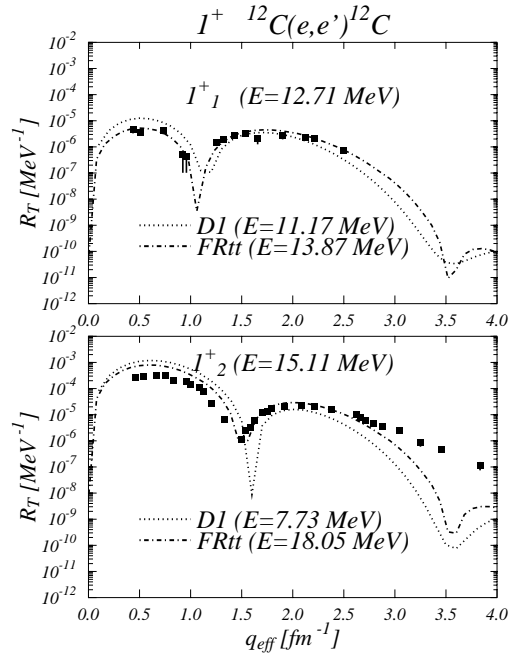}
\end{center}
\vskip -1cm
\caption{Electron scattering transverse responses 
of the first $1^+$ states of \car, versus the effective momentum
transfer. The self-consistent results obtained with the
Gogny D1 interaction are compared with those corresponding to the FRtt
phenomenological case.  
Experimental data from~\cite{but86,hic84}.}
\label{fig:c12}
\end{figure}

The transverse response functions of the $12^-$ states of \pb are
shown in the right panels of
figure~\ref{fig:pb12m}. We can see that for the higher state
at 7.08 MeV (lower right panel) 
the experimental data are rather well reproduced with
all NN interactions, which do
not produce significant
differences
in the curves. On the other hand, for the lower energy state (upper
right panel)
we observe a very strong dependence on the residual interaction, both
when finite-range and when tensor channel contributions are included.
We remark that a better description of this state alone could be
obtained
with a different choice of the parameters of the residual interaction,
but this would worsen the global descritpion of the various magnetic spectra
we have considered.

In the left panels of the same figure 
the transverse responses of the $10^-$ states
are also shown. It is interesting
to notice that, in this case, only the curves which include tensor
contributions are able to reproduce the second peak shown by the data,
for both states.

An example of the results we have obtained  within 
the self-consistent approach is
given in table~\ref{tab:c12}, where the  low energy magnetic 
spectrum of \car
is presented, and in figure~\ref{fig:c12}, where we show the 
corresponding transverse
responses of the $1^+$ isospin-doublet of states.
For the sake of comparison between the self-consistent results and the purely
phenomenological ones, we have taken as reference the
FRtt case, considered to be the 
most complete interaction we have obtained.

\begin{figure}[t]
\begin{center}
\includegraphics[width=3.5in]{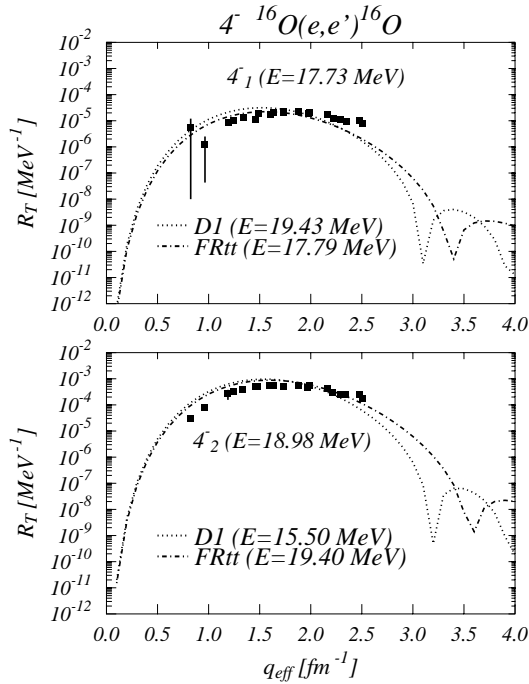}
\end{center}
\vskip -1cm
\caption{Same as figure~\ref{fig:c12}  
for the first $4^-$ states of \oxy.  
Experimental data from~\cite{wri84}.}
\label{fig:o16}
\end{figure}

We see that the magnetic states of \car are reproduced rather badly
by the self-consistent calculations: the response functions agree
in shape and magnitude with the data, but  
the order of the states forming isospin doublets is inverted.

This happens also for \oxy, as illustrated, for the case of the $4^-$
states,
in figure~\ref{fig:o16}, where the values of the excitation energies
obtained with the D1 and FRtt interactions are also reported.

We have systematically obtained this kind of inversion  for all 
magnetic states of all
the nuclei we have studied . This result indicates the inadequacy of
the D1 interaction in isospin-dependent channels, and it may be considered
the most important outcome of the self-consistent  part of our
study.

%
%
\begin{figure}[t]
\begin{center}
\includegraphics[width=5.4in]{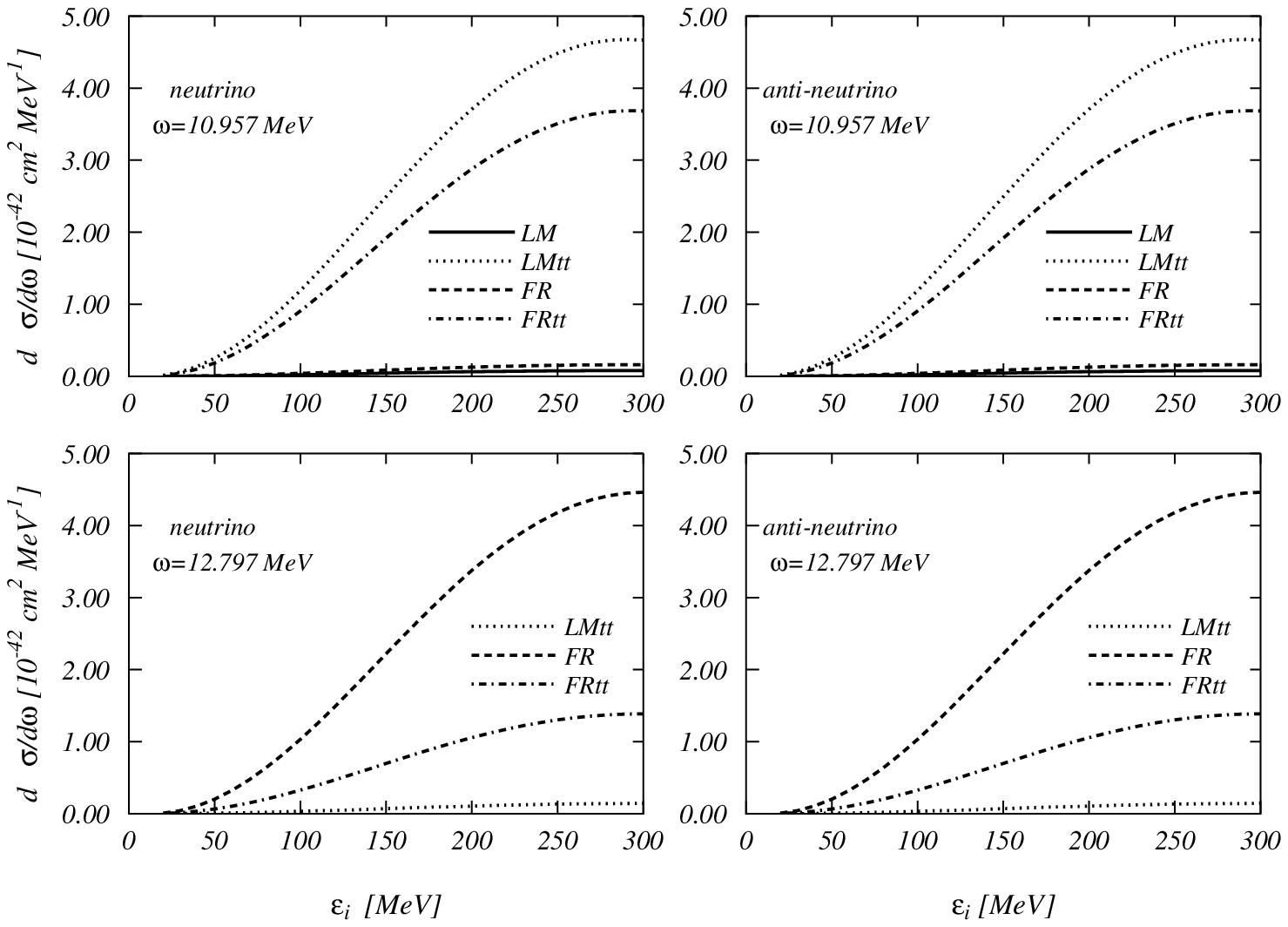}
\end{center}
\vskip -3.2cm
\caption{Cross sections for the neutral current neutrino and
  antineutrino
excitation of $0^-$ states of \oxy at excitation energy $\omega$,
versus the incident (anti)neutrino energy.} 
\label{fig:nu}
\end{figure}
\section{Extension of the phenomenological approach to neutrino
  scattering}

After obtaining an ``optimal'' phenomenological RPA approach, by 
tuning NN interactions to reproduce
data from electron scattering at best, we have also considered the
possibility of applying it to the study of low energy neutrino
scattering cross-sections, with the purpose of studying their
sensitivity to the tensor components of the interaction.

In figure~\ref{fig:nu} we show the cross sections for
the neutral-current neutrino and antineutrino excitation 
of the low energy $0^-$ states
of \oxy, as a function of the incident neutrino energy. The values
of the excitation
energies $\omega$ are indicated inside the figure. 

We observe that, in this test-case, extremely large differences 
in the cross sections are
obtained when tensor channel contributions are included in the
effective NN interaction. This indicates that the role of the residual
interaction in neutrino scattering cross section is a very interesting
topic, for example in connection with the 
problem of nuclear uncertainties in the detection of supernova
neutrinos
\cite{nowproc}.

\section{Summary and conclusions}

We have made a systematic study of the low-lying
(magnetic) spectra of doubly-closed-shell nuclei.
Our study indicates that a simultaneous
description of all states imposes strong constraints on the residual
NN interaction. Within a purely phenomenological approach it is
possible to get a good description of the spectra and of the
response functions of most of
the states, thus obtaining an ``optimal'' RPA approach. Some states
which are not well described exhibit a strong sensitivity to some
details of the residual interaction, and a deeper investigation
could be used to obtain further constraints on it.

On the contrary, the self-consistent calculations we have performed
with
the Gogny D1 interaction produce results in disagreement with the
experimental spectra, indicating the presence of problems in
particular
in the isospin dependent parts of the interaction.  
A detailed study can be found in ~\cite{don09}.

Finally we have considered an example of a potentially important 
application of our phenomenological approach
to the calculation of neutrino scattering cross sections, to
investigate the sensitivity of the latter to the tensor components
of the residual interaction.

\providecommand{\newblock}{}

\end{document}